\documentclass[
floatfix, 
twocolumn,
showpacs, 
showkeys, 
preprintnumbers, 
nofootinbib, 
superscriptaddress
]{revtex4-1}
\usepackage[utf8]{inputenc}
\usepackage[sort&compress]{natbib}
\usepackage{ulem}
\usepackage{bm}
\usepackage{times}
\usepackage{amssymb,amsbsy,amsmath,amsfonts}
\usepackage{graphicx}
\usepackage{float}
\usepackage{color}
\usepackage{morefloats}
\usepackage{rotating}
\usepackage{srcltx}
\usepackage{slashed}
\usepackage{subfigure}
\usepackage{multirow}
\usepackage{verbatim}
\usepackage{hyperref}
\usepackage{tabularx}
\usepackage{braket}
\usepackage{lipsum}

\def\p{\partial}
\def\d{\mathrm{d}}
\def\J{\mathcal{J}}

\def\D{\mathcal{D}} 

\DeclareUnicodeCharacter{3000}{HEREHEREHERE}

\begin{document}

\title{Long-range $S$-wave $DD^*$ interaction in covariant chiral effective field theory }

\author{Qing-Yu Zhai}
\affiliation{School of Physics, Beihang University, Beijing 102206, China}

\author{Ming-Zhu Liu}
\affiliation{ School of Nuclear Science and Technology, Lanzhou University, Lanzhou 730000, China}

\author{Jun-Xu Lu}
\email[Corresponding author: ]{ljxwohool@buaa.edu.cn}
\affiliation{School of Physics, Beihang University, Beijing 102206, China}

\author{Li-Sheng Geng}
\email[Corresponding author: ]{lisheng.geng@buaa.edu.cn}
\affiliation{School of
Physics,  Beihang University, Beijing 102206, China}
\affiliation{Peng Huanwu Collaborative Center for Research and Education, Beihang University, Beijing 100191, China}
\affiliation{Beijing Key Laboratory of Advanced Nuclear Materials and Physics, Beihang University, Beijing 102206, China }
\affiliation{Southern Center for Nuclear-Science Theory (SCNT), Institute of Modern Physics, Chinese Academy of Sciences, Huizhou 516000, Guangdong Province, China}

\begin{abstract}
Motivated by the recent lattice QCD study of the $DD^*$ interaction at unphysical quark masses, we perform a theoretical study of the $DD^*$ interaction in covariant chiral effective field theory (ChEFT). In particular, we calculate the relevant leading-order two-pion exchange contributions. The results compare favorably with the lattice QCD results, supporting the conclusion that the intermediate-range $DD^*$ interaction is dominated by two-pion exchanges and the one-pion exchange contribution is absent. At a quantitative level, the covariant ChPT results agree better with the lattice QCD results than their non-relativistic counterparts, showing the relevance of relativistic corrections in the charm sector.

\end{abstract}

\maketitle

\section{Introduction}
In 2021 the LHCb Collaboration observed a narrow structure in the $D^0D^0\pi^+$ invariant mass spectrum of the $pp$ interaction,  which is identified as a doubly charmed tetraquark state, i.e., $T^+_{cc}(3875)$. Adopting the Breit-Wigner parametrization, its mass and width were determined to be~\cite{LHCb:2021vvq}
\begin{eqnarray}
    \delta m_{\mathrm{BW}} &=& -273 \pm 61 \pm 5 ^{+11}_{-14} ~ {\rm keV},\notag \\
    \Gamma_{\mathrm{BW}} &=& 410 \pm 165 \pm 43 ^{+18}_{-38} ~ {\rm keV}.\notag
\end{eqnarray}
However, it is natural to expect that the width of the $T_{cc}^+$ should be smaller than that of the $D^{*+}$~\cite{Meng:2021jnw,Ling:2021bir,Feijoo:2021ppq,Yan:2021wdl,Fleming:2021wmk,Dai:2023mxm}. Later, the LHCb Collaboration analyzed their data with a resonance profile more suitable to account for the closeness of the $T_{cc}^+$ to the $D^{*+}D^0$ threshold, and the mass and width of $T_{cc}^+$ were found to be~\cite{LHCb:2021auc}
\begin{eqnarray}
    \delta m_{\mathrm{pole}} &=& -360 \pm 40 ^{+4}_{-0} ~ {\rm keV},\notag \\
    \Gamma_{\mathrm{pole}} &=& 48 \pm 2 ^{+0}_{-14} ~ {\rm keV}.\notag
\end{eqnarray}

There were many theoretical studies predicting the existence of a $cc\bar{u}\bar{d}$ tetraquark before the experimental discovery~\cite{Carlson:1987hh,Silvestre-Brac:1993zem,Semay:1994ht,Moinester:1995fk,Pepin:1996id,Gelman:2002wf,Vijande:2003ki,Janc:2004qn,Vijande:2007rf,Lee:2009rt,Yang:2009zzp,Li:2012ss,Feng:2013kea,Karliner:2017qjm,Luo:2017eub,Wang:2017uld,Xu:2017tsr,Junnarkar:2018twb,Liu:2019stu,Qin:2020zlg}. The theoretical
predictions for the mass of the $cc\bar{u}\bar{d}$ ground state with spin-parity quantum numbers $J^P = 1^+$ and isospin $I = 0$, relative to the $D^{*+}D^0$ mass threshold
$\delta m=m_{T_{cc}^+}-(m_{D^0}+m_{D^{*+}})$,
lies in the range $-300 < \delta m < 300$\ MeV. After the LHCb discovery, more studies were performed and some of the earlier studies were updated. As the $T_{cc}^+$ state has very small binding energy and narrow width, the molecular picture has gained a lot of attention~\cite{Chen:2021cfl,Ren:2021dsi,Albaladejo:2021vln,Dong:2021bvy,Du:2021zzh,Xin:2021wcr,Ke:2021rxd,Cheng:2022qcm,Agaev:2022ast,Dai:2023cyo}. In Ref.~\cite{Ling:2021bir}, the effective Lagrangian approach was used to investigate the decay width of $T_{cc}^+$, and the results support its molecular nature. In Ref.~\cite{Cheng:2022qcm}, the one-boson exchange (OBE) potential model and the complex scaling method (CSM) are used and $T_{cc}^+$ is shown to correspond to a quasibound state. In Ref.~\cite{Dai:2023cyo}, the authors found that the possible contribution of a non-molecular component or missing channels is smaller than 3\%, which supported the molecular nature of $T_{cc}^+$.  In addition, there also exist coupled-channel studies ~\cite{Feijoo:2021ppq,Du:2021zzh,Albaladejo:2021vln,Ortega:2022efc,Meng:2021jnw}. In Refs.~\cite{Feijoo:2021ppq,Du:2021zzh}, the authors studied the $T_{cc}^+$ in the $D^{*+}D^0$ and $D^{*0}D^+$ coupled channels, and found a bound state corresponding to the $T_{cc}^+$, while in Ref.~\cite{Ortega:2022efc}, $D^{*+}D^0$, $D^{*0}D^+$, and $D^{*0}D^{*+}$ coupled channels and the constituent quark model were used and it was found that the $D^{*+}D^0$ component accounts for 86\% of the $T_{cc}^+$ wave function. In addition to its production mechanisms and decay properties, its electromagnetic properties~\cite{Deng:2021gnb,Ozdem:2021hmk}, the effects of three-body $DD\pi$ cut~\cite{Du:2021zzh,Du:2023hlu}, the compositeness~\cite{Kinugawa:2023fbf}, and even the yield of $T_{cc}^+$ in heavy ion collisions~\cite{Hu:2021gdg} have been studied. 


After the experimental discovery, several lattice QCD studies have been performed~\cite{Padmanath:2022cvl,Chen:2022vpo,Lyu:2023xro}, In Ref.~\cite{Padmanath:2022cvl}, a simulation of $DD^*$ scattering for $m_\pi\simeq280$ MeV was performed, and a doubly charm tetraquark with $J^P = 1^+$ features as a virtual bound state in the simulation with a charm quark mass slightly larger than its physical value. In Ref.~\cite{Chen:2022vpo}, the S-wave $DD^*$ scattering in the isospin $I = 0,1$ channels was studied for $m_\pi\simeq350$ MeV, and the authors found that the $DD^*$ interaction in the $I = 0$ channel is attractive for a wide range of the $DD^*$ energy, while the $DD^*$ interaction induced by the charged $\rho$ meson exchange may play a crucial role in the formation of $T_{cc}^+(3875)$.

It is particularly interesting to note that in Ref.~\cite{Lyu:2023xro}, the $DD^*$ interaction in the isoscalar and $S$-wave channel is studied for a nearly physical pion mass $m_\pi\simeq146$ MeV, and the long-range part of the potential is found to be dominated by the two-pion exchange at least in the range $1 < r < 2$ fm~\footnote{Note that in Ref.~\cite{Wang:2022jop}, $1<r<2$ fm is referred to as intermediate range. In this work, following Ref.~\cite{Lyu:2023xro}, we refer to $1<r<2$ fm as long range. } while the one-pion exchange potential is absent. The overall attraction is found to be strong enough to generate a near-threshold virtual state, which evolves into a loosely bound state for the physical $m_\pi\simeq135$ MeV.

In Refs.~\cite{Xu:2017tsr,Wang:2022jop}, the one-pion and two-pion exchange potentials in the $DD^*$ system were calculated in the non-relativistic chiral effective field theory (ChEFT) up to the second and third order, respectively. It is interesting to note that in Ref.~\cite{Wang:2022jop}, the non-relativistic ChEFT supports the dominance of the $ae^{-2m_\pi r}/r^2$ behavior of the two-pion exchange, similar to the lattice simulations but for ranges longer than $1 < r < 2$ fm. It was further pointed out that the $ae^{-2m_\pi r}/r^n$ behavior with $n > 2$ may also play a relevant role. 

Motivated by the lattice QCD discovery of the dominant two-pion exchange potential~\cite{Lyu:2023xro} and the discrepancy between the non-relativistic ChEFT and the lattice QCD simulations~\cite{Wang:2022jop}, we adopt the covariant ChEFT to calculate the TPE contributions to the $DD^*$ interaction. Compared to its non-relativistic counterpart, covariant ChEFT not only satisfies all the symmetry constraints but also converges relatively faster. This is shown to be the case for baryon masses~\cite{Ren:2012aj}, magnetic moments~\cite{Geng:2008mf,Xiao:2018rvd}, meson-baryon scattering~\cite{Alarcon:2011zs,Chen:2012nx,Lu:2018zof,Lu:2022hwm}, nucleon-nucleon scattering~\cite{Ren:2016jna,Xiao:2018jot,Ren:2017yvw,Wang:2020myr,Lu:2021gsb}, hyperon-nucleon scattering~\cite{Li:2016paq,Li:2016mln,Li:2018tbt,Song:2018qqm,Liu:2020uxi,Song:2021yab}, the $\Lambda_{c} N$ system~\cite{Song:2020isu,Song:2021war}, and the singly charmed meson sector~\cite{Yao:2015qia,Du:2016tgp,Du:2017ttu}. 
However, covariant ChEFT has not been applied to study the systems of two charmed hadrons such as $D D^{(*)}(\bar{D}^{(*)})$ or $\Sigma_c^{(*)} \bar{D}^{(*)}$. The effective potential extracted by the HAL QCD method~\cite{Ishii:2006ec,Ishii:2012ssm,Aoki:2020bew,Aoki:2011gt,Iritani:2018vfn} provides a unique opportunity to investigate how covariance plays its role in such a heavy flavor system. That is, whether the covariant ChEFT can better describe the lattice QCD simulation~\cite{Lyu:2023xro}. 


This work is organized as follows. In Sec.II we briefly explain the ChEFT approach and calculate the relevant Feynman diagrams. Results and discussions are given in Sec.III, followed by a short summary in the last section. The analytical results for the pertinent Feynman diagrams are relegated to Appendix~\ref{AppendixA} and Appendix~\ref{AppendixB}. The heavy-meson approximation and the comparison with the non-relativistic results are given in Appendix~\ref{AppendixC}. 

\section{THEORETICAL FORMALISM}

\subsection{ Effective Lagrangians }

In the framework of ChEFT, we can expand the amplitudes with a small parameter $\epsilon=\mathrm{max}\{p/\Lambda, m/\Lambda\}$, where $p$ is the momentum of the pion, $m$ is the pion mass or the $D-D^*$ mass splitting, and $\Lambda$ is the breakdown scale of chiral symmetry or the mass of $D^{(*)}$ mesons.

In this work, we only consider the one-pion-exchange (OPE) diagram at the leading order (LO) $O(\epsilon^0)$ and the two-pion-exchange (TPE) diagram at the next-to-leading order (NLO) $O(\epsilon^2)$. For this we first spell out the covariant chiral effective Lagrangian describing the interactions between charmed mesons $D/D^*$ and Nambu-Goldstone bosons (NGB), which reads~\cite{Altenbuchinger:2013vwa}
\begin{align}
    \mathcal{L}=&\langle\D_\mu P\D^\mu P\rangle-m_P^2\langle PP^\dagger\rangle-\langle\D_\mu P^{*\nu}\D^\mu P^{*\dagger}_\nu\rangle\notag\\
    &+m_{P^*}^2\langle P^{*\nu}P^{*\dagger}_\nu\rangle
    +ig_{D}\langle P_\mu^*u^\mu P^\dagger-Pu^\mu P_\mu^{*\dagger}\rangle\notag\\
    &+\frac{g_{D^*}}{2}\langle(P_\mu^*u_\alpha\p_\beta P_\nu^{*\dagger}-\p_\beta P_\mu^*u_\alpha P_\nu^{*\dagger})\epsilon^{\mu\nu\alpha\beta}\rangle
    \label{DDsLagrangian},
\end{align}
where $P = (D^0, D^+, D^+_s)$ and $P^*_\mu = (D^{*0}, D^{*+}, D^{*+}_s)_\mu$, the axial current is $u_\mu=i(\xi^\dagger\p_\mu\xi-\xi\p_\mu\xi^\dagger)$, and the chiral covariant derivative is
\begin{align}
    \D_\mu P_a=\p_\mu P_a-\Gamma_\mu^{ba}P_b\ ,\ \ \D^\mu P_a^\dagger=\p^\mu P_a^\dagger+\Gamma_{ab}^\mu P_b^\dagger,
\end{align}
where $\Gamma_\mu=\frac{1}{2}(\xi^\dagger\p_\mu\xi+\xi\p_\mu\xi^\dagger)$ is the vector current. In the currents, $\xi^2=\exp(i\Phi/f)$ with $f=0.092$\ GeV being the NGB decay constant in the chiral limit and $\Phi$ collecting the octet of NGB fields: 
\begin{align}
    \Phi=\sqrt{2}\begin{pmatrix}
        \frac{\pi^0}{\sqrt{2}}+\frac{\eta}{\sqrt{6}} & \pi^+ & K^+ \\
        \pi^- & -\frac{\pi^0}{\sqrt{2}}+\frac{\eta}{\sqrt{6}} & K^0 \\
        K^- & \bar{K}^0 & -\frac{2}{\sqrt{6}}\eta 
    \end{pmatrix}.
\end{align}

The coupling $g_D$ is determined from the decay width of the $D^{*+}$ and the coupling $g_{D^*}$ is related to $g_{D}$ through the heavy-quark spin symmetry. The values of $g_D$ and $g_{D^*}$ are $1.177$\ GeV and $g_D/m_{D^*}=0.583$, respectively~\cite{Altenbuchinger:2013vwa,Ling:2021lmq}. 

\subsection{ Effective potentials of $DD^{*}$ System }

\begin{figure}[hptb]
    \centering
    \includegraphics[width=4.25cm]{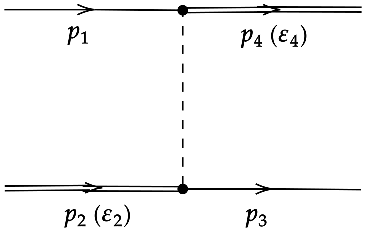}
    \caption{One-pion exchange diagram at LO. The solid, double-solid, and dashed lines stand for $D$, $D^*$, and the pion, respectively. }
    \label{OPE diagram}
\end{figure}

At LO, the OPE diagram is illustrated in Fig.~\ref{OPE diagram}, which mainly contributes to the longest-range interaction. The OPE potential is
\begin{align}
    V_{\mathrm{OPE}}=A_{\mathrm{OPE}}^{I}\cdot\frac{g_D^2}{f^2}\cdot\frac{(\epsilon_{2}\cdot q)(\epsilon_4^{\dagger}\cdot q)}{q^2-m_\pi^{2}+i\epsilon}
    \label{OPE potential}
\end{align}
where $q=p_1-p_4=p_3-p_2$, and $\epsilon_2$ ($\epsilon_4$) is the polarization vector of the $D^*$ meson, and $A_{\mathrm{OPE}}^{I}$ is the isospin factor, where the superscript $I=0,1$ denotes the isospin of the $DD^*$ system. 

\begin{figure}[hptb]
    \centering
    \includegraphics[width=8.5cm]{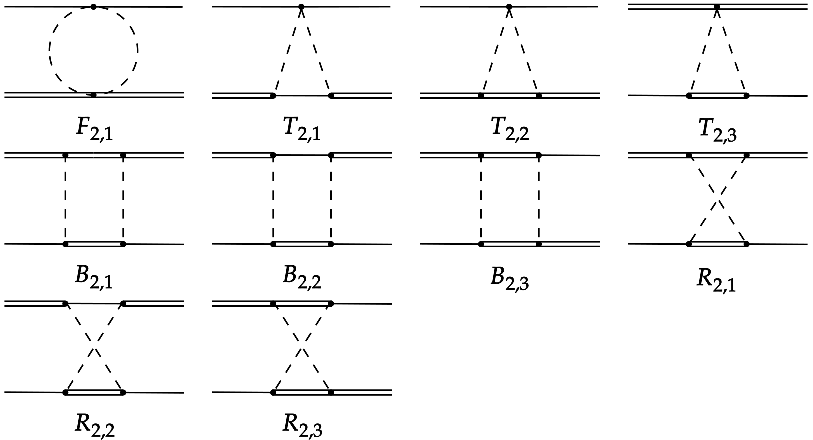}
    \caption{Two-pion exchange diagrams at NLO. The solid, double-solid, and dashed lines stand for $D$, $D^*$, and the pion, respectively.}
    \label{TPE diagram}
\end{figure}

At NLO, there are ten TPE diagrams, which we illustrate in Fig.~\ref{TPE diagram}. With the Lagrangian given in Eq.~(\ref{DDsLagrangian}), one can calculate the TPE potential. Although the procedure is straightforward, the results are a bit tedious. Therefore, we relegate them to the Appendix~\ref{AppendixA}. In Table~\ref{isospin factors} we list the isospin factors appearing in Eq.~(\ref{OPE potential}) and Eq.~(\ref{F21})-Eq.~(\ref{R23}). 

\begin{table}[htpb]
    \centering
    \caption{Isospin factors for the OPE and TPE potentials of $DD^*\rightarrow DD^*$.}\label{tab:isospin factors}
    \begin{tabular}{ c c c c c c c c c }
        \hline\hline
        & $A_{\mathrm{OPE}}^I$ & $A_{F_{2,1}}^I$ & $A_{T_{2,1}}^I$ & $A_{T_{2,2}}^I$ & $A_{T_{2,3}}^I$ & $A_{B_{2,1}}^I$ & $A_{B_{2,2}}^I$ & $A_{B_{2,3}}^I$ \\
        \hline
        $I=0$ & 3 & 3 & 3 & -3 & -3 & -9 & 9 & 9 \\
        $I=1$ & 1 & -1 & -1 & 1 & 1 & -1 & 1 & -1 \\
        \hline\hline
        & $A_{R_{2,1}}^I$ & $A_{R_{2,2}}^I$ & $A_{R_{2,3}}^I$ \\
        \hline
        $I=0$ & 3 & -3 & -3\\
        $I=1$ & -5 & 5 & -5\\
        \hline\hline
    \end{tabular}
    \label{isospin factors}
\end{table}

We calculate the effective potential in the center of mass system (c.m.s) of the $DD^*$ and project the potential to the $S$-wave~\cite{Golak:2009ri}
\begin{align}
    V\big({ }^{2S+1}L_{J}={}^{3}S_{1}\big)=\frac{1}{2}\int{\d^3p'\ V(\pmb{p}, \pmb{p'})\sin{\theta}}
    \label{partial wave}
\end{align}
where $\theta$ is the angle between $\pmb{p}$ and $\pmb{p'}$. We note that the Feynman diagrams $R_{2,1}$ and $R_{2,2}$ have a left-hand cut for $\theta\in[0,\pi]$, and therefore we approximate $m_D=m_{D^*}$ in the dynamics but maintain the threshold of $D$ and $D^*$, i.e., $m_D+m_{D^*}$, the same as the lattice simulations in the kinematics. According to the covariant power-counting rule based on naive dimensional analysis~\cite{Altenbuchinger:2013vwa}, $(m_{D^*}-m_D)$ is of order $\mathcal{O}(q^2)$.Thus taking $(m_{D^*}-m_D)/\Lambda=0$ in the TPE potential is a reasonable approximation and the difference is of higher chiral order which can be neglected. 

We note that most of the TPE diagrams contain ultraviolet divergences which should be absorbed by the corresponding contact interactions of the same order. This, however, requires the introduction of unknown low-energy constants (LECs). In Ref.~\cite{Xu:2017tsr}, these LECs were determined in the resonance saturation approach. In the present work, as our main purpose is to check whether for $1\le r \le 2$ fm, the $DD^*$ interaction is dominated by the TPE contribution, we take a different regularization approach which is physically more intuitive. That is,  we multiply each pion propagator with a monopole form factor
\begin{align}
    F(q^2)=\frac{m_\pi^2-\Lambda^2}{q^2-\Lambda^2},
    \label{Form Factor}
\end{align}
as is usually done in OBE models (see, e.g., Ref.~\cite{Cheng:2022qcm}). We have checked that all the divergences originating from the dimensional regularization can be removed by $F(q^2)$, and the residual part can be considered as the genuine two-pion exchange potential. Note that our covariant chiral TPE potential has dimension $[E]^0$, while the non-relativistic ChEFT potential and the potential obtained in the lattice QCD simulation have dimension $[E]^{-2}$, thus we divide our potential by $\sqrt{2m_D2m_{D^*}2m_D2m_{D^*}}$ to set both dimensions the same~\cite{Sun:2011uh}. 

\subsection{ Subtraction of the reducible part of the TPE potential }

Note that the $DD^*$ interaction extracted from lattice QCD simulations corresponds to the effective potential. As a result, we need to subtract from the amplitude of diagram $B_{2,2}$ the reducible part. Otherwise, there will be double counting when the effective potential is inserted into the Kadyshevsky equation. The reducible part can be calculated in the following way
\begin{align}
    V_{\mathrm{RP}}&=i\int\frac{\d^4l}{(2\pi)^4}\cdot\frac{V_{\mathrm{OPE}}(p,l)}{k_1^2-m_1^2+i\epsilon}\cdot\frac{V_{\mathrm{OPE}}(l,p')}{k_2^2-m_2^2+i\epsilon}\label{iterated OPE}\\
    &\simeq\int\frac{dl}{(2\pi)^3}\cdot\frac{l^2}{4E_1E_2}\frac{V_{\mathrm{OPE}}(p,l)\cdot V_{\mathrm{OPE}}(l,p')}{\sqrt{s}-E_T+i\epsilon}
    \label{iterated OPE final}
\end{align}
where
\begin{align}
    k_1&=\Big(\frac{s-m_{D^*}^2+m_D^2}{2\sqrt{s}}+l_0,\ \pmb{l}\Big), \notag\\
    k_2&=\Big(\frac{s+m_{D^*}^2-m_D^2}{2\sqrt{s}}-l_0,\ -\pmb{l}\Big), \notag
\end{align}
and $E_1=\sqrt{\pmb{l}^2+m_D^2}$, $E_2=\sqrt{\pmb{l}^2+m_{D^*}^2}$, $E_T=E_1+E_2$.

\begin{figure}[hptb]
    \centering
    \includegraphics[width=5.0cm]{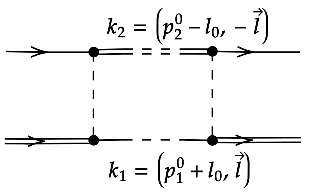}
    \caption{The reducible part of the TPE potential from Feynmann diagram $B_{2,2}$, where $p_1^0=s-m_{D^*}^2+m_D^2/2\sqrt{s}$ and $p_2^0=s+m_{D^*}^2-m_D^2/2\sqrt{s}$. }
\end{figure}

To calculate the integral of Eq.~(\ref{iterated OPE}), we let the momentum of $D^*$ in $V_{\mathrm{OPE}}$ on-shell and the momentum of $D$ off-shell, and close the $l_0$ contour integral in the lower half-plane with the pole located at $l_0^{(1)}=E_1-(s-m_{D^*}^2+m_D^2)/2\sqrt{s}-i\epsilon$ and $l_0^{(2)}=E_2+(s+m_{D^*}^2-m_D^2)/2\sqrt{s}-i\epsilon$. Then Eq.~(\ref{iterated OPE final}) can be easily obtained by using the residue theorem. The detailed calculation is given in Appendix~\ref{AppendixA}. 

\subsection{Lattice QCD $DD^*$ potential in momentum space }
The lattice QCD potential $V_{\mathrm{TPE}}^{\mathrm{L}}(r)$ is given in coordinate space ~\cite{Lyu:2023xro},
\begin{align}
    V_{\mathrm{TPE}}^{\mathrm{L}}(r)=a_3\frac{\mathrm{e}^{-2m_\pi r}}{r^2}\label{VL in coodinate space},
\end{align}
where $a_3=-0.045$\ GeV is the fitted parameter in lattice simulation. 
To compare with the potential obtained in covariant ChEFT, we need to transform it into momentum space
\begin{align}
    V_{\mathrm{TPE}}^{\mathrm{L}}(\pmb{q})=\int\d^3\pmb{q}\ V_{\mathrm{TPE}}^{\mathrm{L}}(r)e^{-i\pmb{q}\cdot\pmb{r}}.
    \label{VL in coodinate space}
\end{align}
Performing the integration analytically, we obtain
\begin{align}
   V_{\mathrm{TPE}}^{\mathrm{L}}(q)=\tilde{a}_3\ \frac{4\pi}{q}\arctan\Big(\frac{q}{2m_\pi}\Big)\label{VL in momentum space}
\end{align}
where $\tilde{a}_3\simeq-1.15$\ GeV$^{-1}$ is the fitted parameter in natural units, and $q=|\pmb{q}|=\sqrt{p^2+p'^2-2pp'\cos\theta}$, where $\pmb{q}$ is the transfer momentum. Note that the potential of  Eq.~(\ref{VL in momentum space}) should be projected to the $S$-wave before it can be compared with both the relativistic and non-relativistic chiral potentials. 

\section{ Numerical Results and Discussions }

In this section, for the sake of convenience, we use $V_{\mathrm{OPE}}^{\mathrm{R}}$ and $V_{\mathrm{TPE}}^{\mathrm{R}}$ to denote our relativistic OPE and TPE potentials, respectively. $V_{\mathrm{TPE}}^{\mathrm{NR}}$ refers to the non-relativistic TPE potential in momentum space derived in Ref.~\cite{Wang:2022jop} and $V_{\mathrm{TPE}}^{\mathrm{L}}$ refers to the TPE potential given in Eq.~(\ref{VL in momentum space}) Fourier transformed from the lattice QCD potential~\cite{Lyu:2023xro}. 

We first compare our covariant chiral potentials with those obtained in the heavy meson chiral effective field theory (HMChEFT)~\cite{Xu:2017tsr}. The details of the analytic results are shown in the Appendix~\ref{AppendixC}, where at the non-relativistic limit, i.e., $p\rightarrow 0$  and $m_{D^{(*)}}\rightarrow\infty$, our potentials have the same analytic structure as those of HMChEFT, as they should be. However, for finite momentum and heavy quark masses, they will be different, as we will see later. 

\begin{figure}[htpb]
    \centering
    \includegraphics[width=8.5cm]{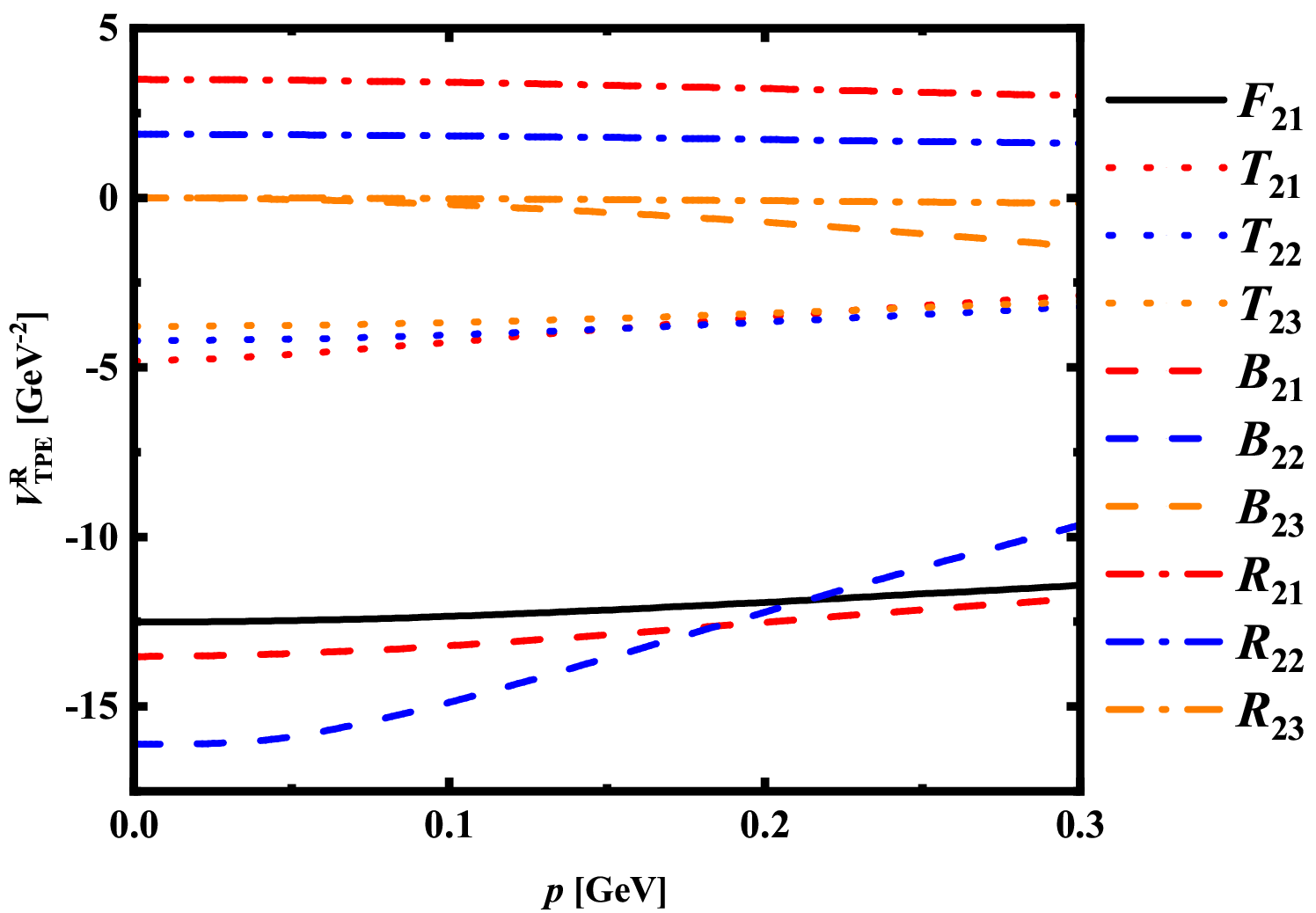}
    \caption{Contributions of different Feynman diagrams as a function of  $p=\sqrt{[s-(m_D+m_{D^*})^2][s-(m_D-m_{D^*})^2]}/2\sqrt{s}$ in the c.m.s. of $DD^*$ for a cutoff of $\Lambda\simeq0.95$\ GeV. 
    }
    \label{TPE with FF}
\end{figure}

In Fig.~\ref{TPE with FF} we show the contributions of various diagrams for a cutoff of $\Lambda \simeq 0.95$\ GeV to match $V_{\mathrm{TPE}}^{\mathrm{L}}$ at $p=0$. One can see that the relativistic TPE potential is dominated by diagram $B_{2,2}$ at $p\simeq0$, or in the long range, because the pion mass is close to the $D-D^*$ mass splitting, thus the four propagators of diagram $B_{2,2}$ can approach their on-shell conditions simultaneously and therefore enhance the contribution, which is the so-called ``box singularity"~\cite{Duan:2023dky}. The amplitude of diagram $B_{2,2}$ is also sensitive to the c.m.s. momentum and provides the dominant contribution to the momentum dependence of the TPE potential. 

\begin{figure}[htpb]
    \centering
    \includegraphics[width=8.5cm]{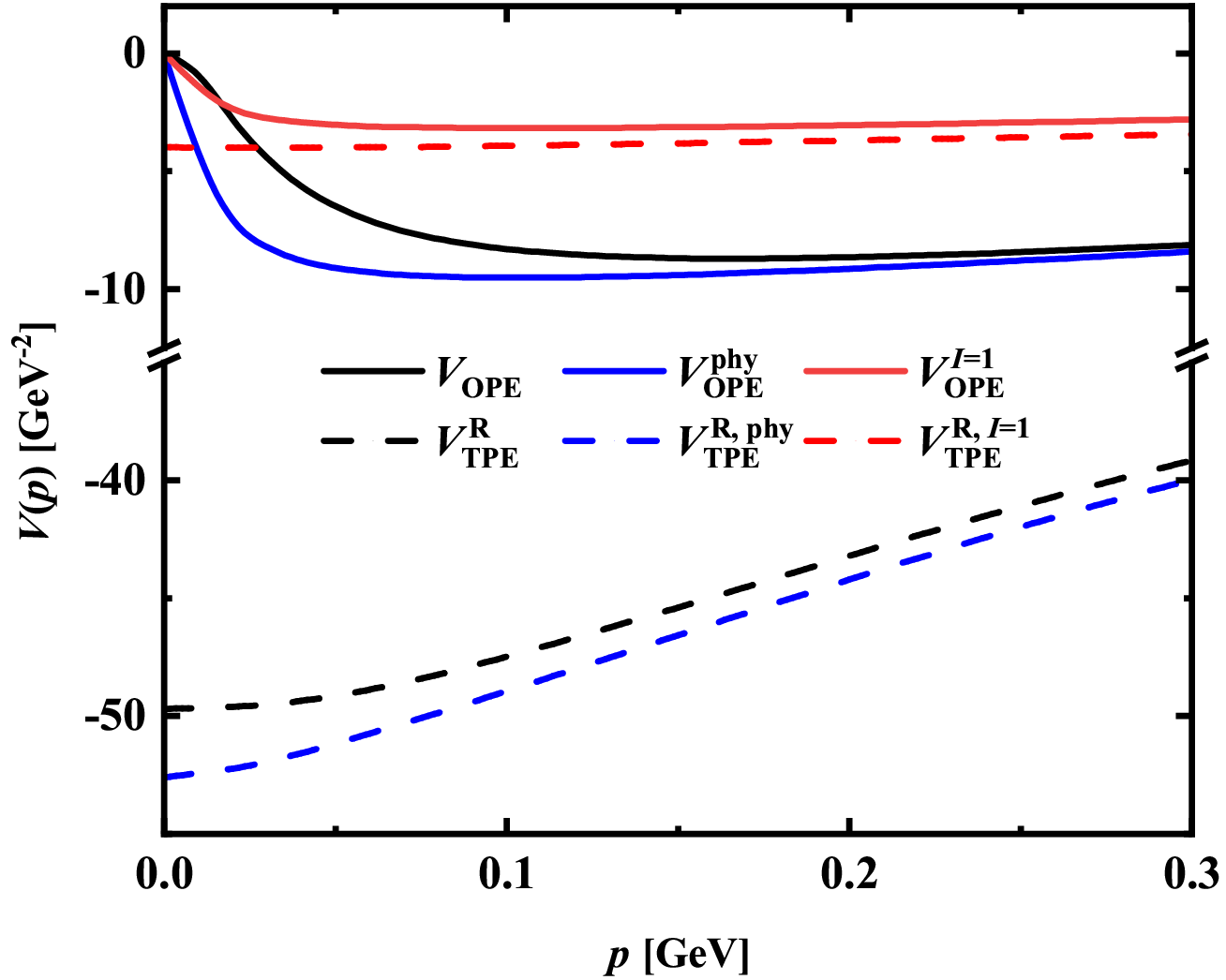}
    \caption{$V_{\mathrm{OPE}}$ and $V_{\mathrm{TPE}}^{\mathrm{R}}$ for different pion masses and different isospin channels for a cutoff $\Lambda\simeq0.95$, where the solid lines denote the OPE potentials and the dashed lines denote the TPE potentials. The black and blue lines denote the potentials for $m_\pi=0.146$\ GeV, $m_D=1.878$\ GeV, $m_{D^*}=2.018$\ GeV and $m_\pi=0.138$\ GeV, $m_D=1.870$\ GeV, $m_{D^*}=2.007$\ GeV. The red lines denote the OPE and TPE potentials in the $I=1$ channel. }
    \label{different potentials}
\end{figure}

Next, in Fig.~\ref{different potentials} we show the OPE and TPE potentials for a cutoff of  $\Lambda=0.95$\ GeV in the $I=0$ channel. One can see that both the OPE and TPE potentials are attractive for all the ranges, and $V_{\mathrm{OPE}}^{\mathrm{R}}\simeq0$ as $p\simeq0$. In particular, $V_{\mathrm{OPE}}$ is much weaker than $V_{\mathrm{TPE}}^{\mathrm{R}}$ in the long range, which is consistent with the observation that $V_{\mathrm{OPE}}$ is absent in the lattice QCD simulation~\cite{Lyu:2023xro}. 

In Fig.~\ref{different potentials} we also show the OPE and TPE potentials for the physical pion mass in the $I=0$ channel. Compared with the potentials obtained with the lattice QCD pion mass, the interaction becomes more attractive. This is also consistent with the lattice QCD study where it was shown that the potential for $m_\pi=0.146$\ GeV produces a virtual state, while a loosely bound state is generated for  $m_\pi=0.135$\ GeV. 

We also show the isovector OPE and TPE potentials in Fig.~\ref{different potentials}. They become comparable to each other in the range of 0.05\ --\ 0.3\ GeV, and although both are attractive, they are less attractive than their isoscalar counterparts. Since the isoscalar OPE and TPE potentials (black lines) can barely produce a virtual (or a loosely bound) state~\cite{Lyu:2023xro}, and the contact interaction determined in the resonance saturation model is repulsive~\cite{Xu:2017tsr}, we can assert that there exists no bound state in the $I=1$ channel, consistent with Refs.~\cite{Xu:2017tsr,Li:2012ss,Ohkoda:2012hv}.

\begin{figure}[htpb]
    \centering
    \includegraphics[width=8.5cm]{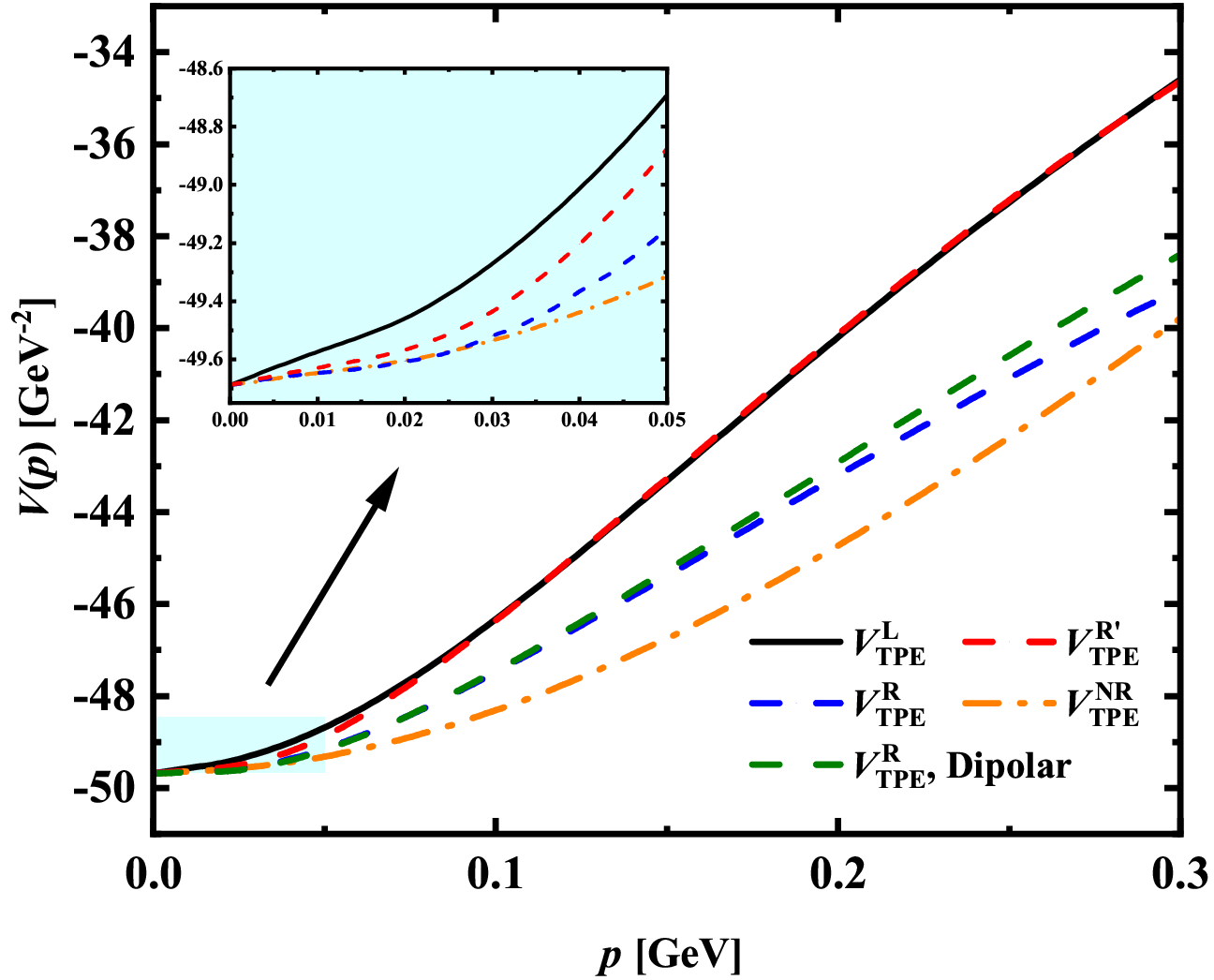}
    \caption{$V_{\mathrm{TPE}}^{\mathrm{L}}$, $V_{\mathrm{TPE}}^{\mathrm{R}}$ and $V_{\mathrm{TPE}}^{\mathrm{NR}}$. We used  a cutoff $\Lambda\simeq0.95$\ GeV for $V_{\mathrm{TPE}}^{\mathrm{L}}$ and $\Lambda\simeq1.4$\ GeV for $V_{\mathrm{TPE}}^{\mathrm{NR}}$ to match $V_{\mathrm{TPE}}^{\mathrm{L}}$ at $p=0$. The $V_{\mathrm{TPE}}^{\mathrm{R'}}$ which is denoted by a red dashed line is the potential we use unphysical parameter $f$ with $\Lambda\simeq0.74$\ GeV. }
    \label{Lattice TPE}
\end{figure}

In Fig.~\ref{Lattice TPE}, we compare $V_{\mathrm{TPE}}^{\mathrm{L}}$, $V_{\mathrm{TPE}}^{\mathrm{R}}$, and $V_{\mathrm{TPE}}^{\mathrm{NR}}$. We set $\Lambda\simeq0.95$\ GeV for $V_{\mathrm{TPE}}^{\mathrm{R}}$ and $\Lambda\simeq1.4$\ GeV for $V_{\mathrm{TPE}}^{\mathrm{NR}}$ to match $V_{\mathrm{TPE}}^{\mathrm{L}}$ at $p=0$. We can see that $V_{\mathrm{TPE}}^{\mathrm{R}}$ is consistent with $V_{\mathrm{TPE}}^{\mathrm{NR}}$ between 0 and 0.03\ GeV, consistent with our analytic results shown in Appendix~\ref{AppendixC}. In addition, $V_{\mathrm{TPE}}^{\mathrm{R}}$ increases faster than $V_{\mathrm{TPE}}^{\mathrm{NR}}$ in the long-range (or in the small-momentum 0\ --\ 0.3\ GeV) region. 

The different behavior between $V_{\mathrm{TPE}}^{\mathrm{NR}}$ and $V_{\mathrm{TPE}}^{\mathrm{R}}$ mainly originates from the non-relativistic approximation, i.e., the neglect of terms of $\mathcal{O}(m_D)$ or higher order. 

The lattice QCD simulations~\cite{Lyu:2023xro} show that in the range of $1<r<2$\ fm the $DD^*$ interaction can be described by a TPE potential, i.e., Eq.~(\ref{VL in coodinate space}) which corresponds to Eq.~(\ref{VL in momentum space}) in momentum space as shown in Fig.~\ref{Lattice TPE}. In Ref.~\cite{Wang:2022jop}, it was shown that $V_{\mathrm{TPE}}^{\mathrm{NR}}$ with a cutoff of $\Lambda\in[0.4, 0.9]$\ GeV has the same asymptotic behavior as the lattice QCD potential, i.e., Eq.~(\ref{VL in coodinate space}), but dominates the range longer than $1<r<2$\ fm. This can be seen in Fig.~\ref{Lattice TPE}\ \textemdash\ because  $1<r<2$\ fm corresponds to $0.07<p<0.15$\ GeV approximately, a longer range corresponds to the momentum smaller than 0.07\ GeV, while in Fig.~\ref{Lattice TPE} we can see that in the range of $0<p<0.03$\ GeV especially, both $V_{\mathrm{TPE}}^{\mathrm{R}}$ and $V_{\mathrm{TPE}}^{\mathrm{NR}}$ are similar to $V_{\mathrm{TPE}}^{\mathrm{L}}$.
In the range of $0<p<0.3$\ GeV, with a reasonable cutoff, our covariant TPE potential can better describe $V_{\mathrm{TPE}}^{\mathrm{L}}$ than the non-relativistic TPE. 

Note that $V_{\mathrm{TPE}}^{\mathrm{R}}$ shown in Fig.~\ref{Lattice TPE} does not agree well with $V_{\mathrm{TPE}}^{\mathrm{L}}$. To understand whether this is due to the particular form factor we used, we tried a dipolar form factor instead of the monopolar one
\begin{align}
    F(q^2)=\Big(\frac{m_\pi^2-\Lambda^2}{q^2-\Lambda^2}\Big)^2\notag
\end{align}
to take into account the internal structure of the pion at each interaction vertex. The results are also shown in Fig.~\ref{Lattice TPE}. We found that the dipolar form factor only improves the agreement between $V_{\mathrm{TPE}}^{\mathrm{R}}$ and $V_{\mathrm{TPE}}^{\mathrm{L}}$ a little bit, thus gives nearly the same description. Meanwhile, this nearly negligible improvement is at the expense of a large cutoff $\Lambda\simeq1.6$\ GeV, which is not very reasonable in the chiral effective field theory. 

On the other hand, we believe that the differences between our $V_{\mathrm{TPE}}^{\mathrm{R}}$ and $V_{\mathrm{TPE}}^{\mathrm{L}}$ can be attributed to the fact that we used physical $g_D$ and $f$ and unphysical meason masses in the derivation of the TPE potential. The quark mass dependence of these quantities needs to be carefully taken into account for a proper comparison with the lattice QCD potential and is highly nontrivial. This goes beyond the scope of the present work. For demonstration, we replace $f'=C_\pi\cdot f^{\mathrm{phy}}$ with $f^{\mathrm{phy}}$. The results are shown in Fig.~\ref{Lattice TPE} for a cutoff $\Lambda\simeq0.74$\ GeV. We find that with this reasonable cutoff our TPE agrees well with $V_{\mathrm{TPE}}^{\mathrm{L}}$ in the whole range of $0.075<p<0.3$\ GeV for $C_\pi\simeq0.9$. Note that in Refs.~\cite{Gasser:1983yg,Beane:2002xf}, the $m_\pi$-dependence of $f_\pi$ up to NLO reads
\begin{align}
    f_\pi=f_\pi^{(0)}\left[1-\frac{m_\pi^2}{4\pi^2(f_\pi^{(0)})^2}\log\left(\frac{m_\pi}{m_\pi^{\mathrm{phy}}}\right)+\cdots\right]\notag
\end{align}
which shows that a large $m_\pi$ corresponds to a small $f_\pi$, thus our use of a smaller $f_\pi$ reasonable. Overall, our results support the conclusion of the lattice QCD simulation that the TPE potential is dominant in the range of $1<r<2$\ fm.

\section{ Summary }
In this work, we have studied the long-range $S$-wave $DD^*$ interaction in covariant chiral effective field theory. In particular, we calculated the one-pion-exchange (OPE) and two-pion-exchange (TPE) potentials, regulated with a molopolar form factor $F(q^2)$. 
We found that the momentum dependence of the TPE potential is dominated by Feynman diagram $B_{2,2}$. By comparing our OPE and TPE potentials, we found that in the $I=0$ channel, the TPE potential is dominant in the long range, and both potentials become more attractive as the pion mass approaches the physical point. While in the $I=1$ channel, the OPE and TPE potentials are comparable and are less attractive than those of the $I=0$ channel. All the results are consistent with the lattice QCD study~\cite{Lyu:2023xro}. 

We compared our covariant TPE potential with the non-relativistic one and the lattice QCD poential, and we found that all three potentials share nearly the same behavior in a range longer than $2$\ fm, while between 1 and 2 fm, our covariant TPE potential describes better the lattice QCD potential. We further demonstrated that with a smaller pion decay constant, e.g., by 10\% compared with its physical value, one can describe better the lattice QCD potential with a reasonable cutoff. Overall, our study supports the conclusion of the lattice QCD study that the two-pion-exchange potential is dominant in the $1<r<2$\ fm. 

In Ref.~\cite{Lyu:2022imf}, it was shown that the long-range $N\Phi$ potential is also dominated by the TPE potential. In Ref.~\cite{Dong:2021lkh}, it was argued that the TPE potential can play a relevant role in the $J/\psi-J/\psi$ interaction. We plan to study these systems in the future in chiral effective field theory. 


\section{Acknowledgments}

Qing-Yu Zhai thanks Dr. Bo Wang for the useful discussions. 
This work is partly supported by the National
Natural Science Foundation of China under Grants Nos.11735003,
11975041, 11961141004, and the fundamental Research
Funds for the Central Universities. Ming-Zhu Liu acknowledges support from the National Natural Science Foundation of China under Grant No.12105007.  Junxu Lu acknowledges support from the National Natural Science Foundation of China under Grant No.12105006. 

\appendix
\begin{widetext}
\section{ Covariant TPE potential }
In this Appendix, we show the explicit expressions of the TPE potentials derived in covariant chiral effective field theory. They read:
\begin{align}	    
    V_{F_{2,1}}&=A_{F_{2,1}}^I\cdot\frac{1}{8f^4}\cdot\frac{1}{2}\int\frac{i\ \d^4l}{(2\pi)^4}\ \frac{[(p_2+p_4)\cdot(p_2-p_4+2l)]\ [(p_1+p_3)\cdot(p_2-p_4+2l)]}{[(p_2-p_4+l)^2-m^2+i\epsilon](l^2-m^2+i\epsilon)}\ (\epsilon_2\cdot\epsilon_4^{\dagger})\label{F21}\\
    V_{T_{2,1}}&=A_{T_{2,1}}^I\cdot\frac{g_{D}^2}{2f^4}\int\frac{-i\ \d^4l}{(2\pi)^4}\ \frac{[(p_4-p_2+l)\cdot\epsilon_4^{\dagger}]\ [(p_1+p_3)\cdot(p_4-p_2+2l)]\ (l\cdot\epsilon_2)}{[(p_2-l)^2-m_D^2+i\epsilon](l^2-m_\pi^2+i\epsilon)[(p_4-p_2+l)^2-m_\pi^2+i\epsilon]}\label{T21}\\
    V_{T_{2,2}}&=A_{T_{2,2}}^I\cdot\frac{g_{D^*}^2}{8f^4}\int\frac{-i\ \d^4l}{(2\pi)^4}\ \epsilon^{\mu\nu\alpha\beta}\epsilon_4^{\dagger\nu}(p_4-p_2+l)^\alpha(p_2+p_4-l)^\beta\ \Big[-g^{\mu\delta}+\frac{(p_2-l)^\mu(p_2-l)^\delta}{m_{D^*}^2}\Big]\ \epsilon^{\lambda\delta\rho\sigma}\epsilon_2^{\lambda}\ l^{\rho}\ (2p_2-l)^{\sigma}\notag\\
    &\qquad\qquad\qquad\frac{(p_1+p_3)\cdot(p_4-p_2+2l)}{[(p_2-l)^2-m_{D^*}^2+i\epsilon](l^2-m_\pi^2+i\epsilon)[(p_4-p_2+l)^2-m_\pi^2+i\epsilon]}\label{T22}\\
    V_{T_{2,3}}&=A_{T_{2,3}}^I\cdot\frac{g_{D}^2}{2f^4}\int\frac{-i\ \d^4l}{(2\pi)^4}\ \frac{[(p_2+p_4)\cdot(p_2-p_4+2l)]\ (\epsilon_2\cdot\epsilon_4^{\dagger})}{[(p_1-l)^2-m_{D^*}^2+i\epsilon](l^2-m_\pi^2+i\epsilon)[(p_2-p_4+l)^2-m_\pi^2+i\epsilon]}\notag\\
    &\qquad\qquad\qquad\ (p_2-p_4+l)^\rho\Big[-g^{\rho\lambda}+\frac{(p_1-l)^\rho(p_1-l)^\lambda}{m_{D^*}^2}\Big]\ l^\lambda\label{T23}\\
    V_{B_{2,1}}&=A_{B_{2,1}}^I\cdot\frac{g_{D}^2g_{D^*}^2}{4f^4}\int\frac{i\ \d^4l}{(2\pi)^4}\ \epsilon^{\mu\nu\alpha\beta}\epsilon_4^{\dagger\nu}(p_4-p_2+l)^\alpha(p_2+p_4-l)^\beta\ \Big[-g^{\mu\lambda}+\frac{(p_2-l)^\mu(p_2-l)^\lambda}{m_{D^*}^2}\Big]\ \epsilon^{\rho\lambda\sigma\eta}\epsilon_2^\rho\ l^\sigma\ (2p_2-l)^\eta\notag\\
    &\qquad\qquad\qquad\qquad\frac{(p_4-p_2+l)^\gamma\Big[-g^{\gamma\delta}+\frac{(p_1+l)^\gamma(p_1+l)^\delta}{m_{D^*}^2}\Big]\ l^\delta}{[(p_1+l)^2-m_{D^*}^2+i\epsilon][(p_2-l)^2-m_{D^*}^2+i\epsilon](l^2-m_\pi^2+i\epsilon)[(p_4-p_2+l)^2-m_\pi^2+i\epsilon]}\label{B21}\\
    V_{B_{2,2}}&=A_{B_{2,2}}^I\cdot\frac{g_{D}^4}{f^4}\int\frac{i\ \d^4l}{(2\pi)^4}\ \frac{[(p_4-p_2+l)\cdot\epsilon_4^{\dagger}]\ (\epsilon_2\cdot l)}{[(p_1+l)^2-m_{D^*}^2+i\epsilon][(p_2-l)^2-m_D^2+i\epsilon](l^2-m_\pi^2+i\epsilon)[(p_4-p_2+l)^2-m_\pi^2+i\epsilon]}\notag\\
    &\qquad\qquad\qquad l^\rho\ \Big[-g^{\rho\lambda}+\frac{(p_1+l)^\rho(p_1+l)^\lambda}{m_{D^*}^2}\Big](p_4-p_2+l)^\lambda\label{B22}\\
    V_{B_{2,3}}&=A_{B_{2,3}}^I\cdot\frac{g_{D}^2g_{D^*}^2}{4f^4}\int\frac{i\ \d^4l}{(2\pi)^4}\ \epsilon^{\rho\sigma\delta\eta}\epsilon_2^{\rho}\ l^\sigma\ (2p_2-l)^\eta\Big[-g^{\delta\gamma}+\frac{(p_2-l)^\delta(p_2-l)^\gamma}{m_{D^*}^2}\Big](p_2-p_3-l)^\gamma\notag\\
    &\qquad\qquad\qquad\qquad\frac{l^\lambda\ \Big[-g^{\lambda\mu}+\frac{(p_1+l)^\lambda(p_1+l)^\mu}{m_{D^*}^2}\Big]\ \epsilon^{\mu\nu\alpha\beta}\epsilon_4^{\dagger\nu}(p_2-p_3-l)^\alpha(p_1+p_4+l)^\beta}{[(p_1+l)^2-m_{D^*}^2+i\epsilon][(p_2-l)^2-m_{D^*}^2+i\epsilon](l^2-m_\pi^2+i\epsilon)[(p_2-p_3-l)^2-m_\pi^2+i\epsilon]}\label{B23}\\
    V_{R_{2,1}}&=A_{R_{2,1}}^I\cdot\frac{g_{D}^2g_{D^*}^2}{4f^4}\int\frac{i\ \d^4l}{(2\pi)^4}\ \epsilon^{\mu\nu\alpha\beta}\epsilon_4^{\dagger\nu}(p_4-p_2+l)^\alpha(p_2+p_4-l)^\beta\Big[-g^{\mu\lambda}+\frac{(p_2-l)^\mu(p_2-l)^\lambda}{m_{D^*}^2}\Big]\epsilon^{\rho\lambda\sigma\eta}\epsilon_2^\rho\ l^\sigma\ (2p_2-l)^\eta\notag\\
    &\qquad\qquad\qquad\qquad\frac{(p_4-p_2+l)^\gamma\Big[-g^{\gamma\delta}+\frac{(p_3-l)^\gamma(p_3-l)^\delta}{m_{D^*}^2}\Big]\ l^\delta}{[(p_3-l)^2-m_{D^*}^2+i\epsilon][(p_2-l)^2-m_{D^*}^2+i\epsilon](l^2-m_\pi^2+i\epsilon)[(p_4-p_2+l)^2-m_\pi^2+i\epsilon]}\label{R21}\\
    V_{R_{2,2}}&=A_{R_{2,2}}^I\cdot\frac{g_{D}^4}{f^4}\int\frac{i\ \d^4l}{(2\pi)^4}\ \frac{[(p_4-p_2+l)\cdot\epsilon_4^{\dagger}]\ (\epsilon_2\cdot l)}{[(p_2-l)^2-m_D^2+i\epsilon][(p_3-l)^2-m_{D^*}^2+i\epsilon](l^2-m_\pi^2+i\epsilon)[(p_4-p_2+l)^2-m_\pi^2+i\epsilon]}\notag\\
    &\qquad\qquad\qquad l^\rho\ \Big[-g^{\rho\lambda}+\frac{(p_3-l)^\rho(p_3-l)^\lambda}{m_{D^*}^2}\Big](p_4-p_2+l)^\lambda\label{R22}\\
    V_{R_{2,3}}&=A_{R_{2,3}}^I\cdot\frac{g_{D}^2g_{D^*}^2}{4f^4}\int\frac{i\ \d^4l}{(2\pi)^4}\ \epsilon^{\rho\sigma\delta\eta}\epsilon_2^{\rho}\ l^\sigma\ (2p_2-l)^\eta\Big[-g^{\delta\gamma}+\frac{(p_2-l)^\delta(p_2-l)^\gamma}{m_{D^*}^2}\Big](p_3-p_2+l)^\gamma\notag\\
    &\qquad\qquad\qquad\qquad\frac{(p_3-p_2+l)^\lambda\Big[-g^{\lambda\mu}+\frac{(p_4-l)^\lambda(p_4-l)^\mu}{m_{D^*}^2}\Big]\ \epsilon^{\mu\nu\alpha\beta}\epsilon_4^{\dagger\nu}\ l^\alpha\ (2p_4-l)^\beta}{[(p_2-l)^2-m_{D^*}^2+i\epsilon][(p_4-l)^2-m_{D^*}^2+i\epsilon](l^2-m_\pi^2+i\epsilon)[(p_3-p_2+l)^2-m_\pi^2+i\epsilon]}\label{R23}
\end{align}
We calculate these loop integrals using dimensional regularization. 
\label{AppendixA}
\end{widetext}

\begin{widetext}
    \section{ Subtraction of the reducible part of Feynman diagram $B_{2,2}$}
    The reducible part of Feynman diagram $B_{2,2}$ is
    \begin{align}
        V_{\mathrm{RP}}&=i\int\frac{\d^4l}{(2\pi)^4}\cdot\frac{V_{\mathrm{OPE}}(p,l)}{k_1^2-m_1^2+i\epsilon}\cdot\frac{V_{\mathrm{OPE}}(l,p')}{k_2^2-m_2^2+i\epsilon}
        \label{iterated OPE_appendix}
    \end{align}
    where
    \begin{align}
        k_1=\Big(\frac{s-m_{D^*}^2+m_D^2}{2\sqrt{s}}+l_0,\ \pmb{l}\Big),\ \ k_2=\Big(\frac{s+m_{D^*}^2-m_D^2}{2\sqrt{s}}-l_0,\ -\pmb{l}\Big) \notag.
    \end{align}
    To calculate Eq.~(\ref{iterated OPE_appendix}), we set $D^*$ on-shell, and $D$ off-shell, and obtain
    \begin{align}
        V_{\mathrm{RP}}\simeq i\int\frac{\d^3l}{(2\pi)^3}V_{\mathrm{OPE}}(p,\sqrt{s}-E_2,E_2,\pmb{l})V_{\mathrm{OPE}}(\sqrt{s}-E_2,E_2,\pmb{l},p')\int\frac{dl_0}{2\pi}\cdot\frac{1}{k_1^2-m_1^2+i\epsilon}\cdot\frac{1}{k_2^2-m_2^2+i\epsilon}
        \label{iterated OPE simplified1_appendix}
    \end{align}
    where $E_2=\sqrt{\pmb{l}^2+m_{D^*}^2}$. The integral of $l_0$ can be calculated by using the residue theorem. We close the $l_0$ contour integral in the lower half-plane with the poles located at
    \begin{align}
        l_0^{(1)}=E_1-\frac{(s-m_{D^*}^2+m_D^2)}{2\sqrt{s}}-i\epsilon\notag\\
        l_0^{(2)}=E_2+\frac{(s+m_{D^*}^2-m_D^2)}{2\sqrt{s}}-i\epsilon\notag
    \end{align}
    where $E_1=\sqrt{\pmb{l}^2+m_D^2}$. Then we have
    \begin{align}
        V_{\mathrm{RP}}\simeq \int\frac{\d^3l}{(2\pi)^3}\ V_{\mathrm{OPE}}(p,\sqrt{s}-E_2,E_2,\pmb{l})V_{\mathrm{OPE}}(\sqrt{s}-E_2,E_2,\pmb{l},p')\cdot\frac{E_1+E_2}{2E_1E_2}\cdot\frac{1}{s-E_T^2}
    \end{align}
    where $E_T=E_1+E_2$. We use the approximation $\sqrt{s}\simeq E_T$ which is the same as reducing the BbS equation to the Kadyshevsky equation. Then we have
    \begin{align}
        V_{\mathrm{RP}}&=\int\frac{dl}{(2\pi)^3}\cdot\frac{l^2}{4E_1E_2}\frac{V_{\mathrm{OPE}}(p,l)\cdot V_{\mathrm{OPE}}(l,p')}{\sqrt{s}-E_T+i\epsilon}\notag\\
        &=\mathcal{P}\int\frac{dl}{(2\pi)^3}\cdot\frac{l^2}{4E_1E_2}\frac{V_{\mathrm{OPE}}(p,l)\cdot V_{\mathrm{OPE}}(l,p')}{\sqrt{s}-E_T}-i\pi\frac{1}{(2\pi)^3}\frac{p_{\mathrm{cm}}}{4\sqrt{s}}V_{\mathrm{OPE}}(p,p_{\mathrm{cm}})V_{\mathrm{OPE}}(p_{\mathrm{cm}},p')
    \end{align}
    where $p_{\mathrm{cm}}=\sqrt{[s-(m_D+m_{D^*})^2][s-(m_D-m_{D^*})^2]}/2\sqrt{s}$ is the momentum in the c.m.s., and $\mathcal{P}$ denotes the principal value of the integral. 

    In principle, the reducible parts of all the three box diagrams $B_{2,1}$, $B_{2,2}$ and $B_{2,3}$ should be subtracted. However, since we do not take into account the coupled channel of  $DD^*\rightarrow D^*D^*$, and the energy region of our interest is below the threshold of $D^*D^*$, that is, the $D^*D^*$ channel does not open, there exists no double counting for the reducible parts of $B_{2,1}$ and $B_{2,3}$. Therefore only the reducible part of diagram $B_{2,2}$ is subtracted.
    \label{AppendixB}
\end{widetext}

\begin{widetext}
    \section{ Non-relativistic approximation of the covariant NLO Potential }
    Following Refs.~\cite{Xu:2017tsr,Wang:2018atz}, one can decompose the potential to combinations of  $J$ functions employing the following tensor decomposition rules,
        \begin{align}
            i\int\frac{\mu^{4-D}\d^Dl}{(2\pi)^D}\ \frac{\{1,l^\mu,l^\mu l^\rho,\cdots\}}{(l^2-m_1^2+i\epsilon)[(q+l)^2-m_2^2+i\epsilon]}\equiv\{\mathcal{J}^F_0,\ q^\mu\mathcal{J}^F_{11},\ q^\mu q^\rho \mathcal{J}^F_{21}+g^{\mu\rho}{J}^F_{22},\cdots\},
            \label{2-point J func}
        \end{align}
        \begin{align}
            &i\int\frac{\mu^{4-D}\d^Dl}{(2\pi)^D}\ \frac{\{1,l^{\mu},l^{\mu}l^{\nu},l^{\mu}l^{\nu}l^{\rho},\cdots\}}{[(+/-)v\cdot l+\omega+i\epsilon](l^2-m_1^2+i\epsilon)[(q+l)^2-m_2^2+i\epsilon]}\notag\\
            \equiv&\{\J_{0}^{T/S},\ q^\mu\J_{11}^{T/S}+v^\mu\J_{12}^{T/S},\ g^{\mu\nu}\J_{21}^{T/S}+q^\mu q^\nu\J_{22}^{T/S}+v^\mu v^\nu\J_{23}^{T/S}+(q\vee v)\J_{24}^{T/S},\notag\\
            &(g\vee q)\J_{31}^{T/S}+q^\mu q^\nu q^\rho\J_{32}^{T/S}+(q^2\vee v)\J_{33}^{T/S}+(g\vee v)\J_{34}^{T/S}+(q\vee v^2)\J_{35}^{T/S}+v^\mu v^\nu v^\rho\J_{36}^{T/S},\ \cdots\},
            \label{3-point J func}
        \end{align}
        \begin{align}
            &i\int\frac{\mu^{4-D}\d^Dl}{(2\pi)^D}\ \frac{\{1,l^{\mu},l^{\mu}l^{\nu},l^{\mu}l^{\nu}l^{\rho},l^{\mu}l^{\nu}l^{\rho}l^{\sigma},\cdots\}}{(v\cdot l+\omega_1+i\epsilon)[(+/-)v\cdot l+\omega_2+i\epsilon](l^2-m_1^2+i\epsilon)[(l+q)^2-m_2^2+i\epsilon]}\notag\\
            \equiv&\{\J_{0}^{R/B},\ q^\mu\J_{11}^{R/B}+v^\mu\J_{12}^{R/B},\ g^{\mu\nu}\J_{21}^{R/B}+q^\mu q^\nu\J_{22}^{R/B}+v^\mu v^\nu\J_{23}^{R/B}+(q\vee v)\J_{24}^{R/B},\notag\\
            &(g\vee q)\J_{31}^{R/B}+q^\mu q^\nu q^\rho\J_{32}^{R/B}+(q^2\vee v)\J_{33}^{R/B}+(g\vee v)\J_{34}^{R/B}+(q\vee v^2)\J_{35}^{R/B}+v^\mu v^\nu v^\rho\J_{36}^{R/B},\notag\\
            &(g\vee g)\J_{41}^{R/B}+(g\vee q^2)\J_{42}^{R/B}+q^{\mu}q^{\nu}q^{\rho}q^{\sigma}\J_{43}^{R/B}+(g\vee v^2)\J_{44}^{R/B}+v^{\mu}v^{\nu}v^{\rho}v^{\sigma}\J_{45}^{R/B}+(q^3\vee v)\J_{46}^{R/B}\notag\\
            &+(q^2\vee v^2)\J_{47}^{R/B}+(q\vee v^3)\J_{48}^{R/B}+(g\vee q\vee v)\J_{49}^{R/B},\ \cdots\}.
            \label{4-point J func}
        \end{align} 
    More specifically, we follow the following steps to perform the non-relativistic reduction of our covariant potentials:
    \begin{enumerate}
    \item Rewrite the denominator\\
    Note that the expression of the $J$ functions can be derived by the heavy meson approach, that is
    \begin{align}
        p_\mu=mv_\mu+k_\mu\ \text{with} \ \ v^2=1. 
    \end{align}
    where $k^2$ is infinitesimal and can be neglected. 
    \item Rewrite the momentum. \\
    We approximate the momentum with the velocity, i.e., 
    \begin{align}
        \frac{p_\mu}{m}\simeq v_\mu
    \end{align}
    and set $m_D=m_{D^*}=m$. In Refs.~\cite{Xu:2017tsr,Wang:2018atz}, the definition of $q$ and $p$ is $q=p_1-p_3$ and $p=p_1-p_4$, we take the following limits
    \begin{align}
        p_2\cdot\epsilon_2\rightarrow0\ &,\ \ p_4\cdot\epsilon_4\rightarrow0\ ,\notag\\
        p_2\cdot\epsilon_4\rightarrow-q\cdot\epsilon_4\ &,\ \ p_1\cdot\epsilon_4\rightarrow p\cdot\epsilon_4\ ,\notag\\
        p_3\cdot\epsilon_2\rightarrow p\cdot\epsilon_2\ &,\ \ p_4\cdot\epsilon_2\rightarrow q\cdot\epsilon_2\ ,\notag\\
        p_1\cdot\epsilon_2\rightarrow q\cdot\epsilon_2+p\cdot\epsilon_2\ &,\ \ p_3\cdot\epsilon_4\rightarrow p\cdot\epsilon_4-q\cdot\epsilon_4.
    \end{align}
    In such limits, $\pmb{p_1}=-\pmb{p_2}=(0,0,\tilde{p})$,\ $\pmb{p_3}=-\pmb{p_4}=(\tilde{p}\sin\theta,0,\tilde{p}\cos\theta)$. Thus $\pmb{q}^2=(\pmb{p_1}-\pmb{p_3})^2=2\tilde{p}^2(1-\cos\theta)$, $\pmb{p}^2=(\pmb{p_1}-\pmb{p_4})^2=2\tilde{p}^2(1+\cos\theta)$. 
    \item Calculate the coefficient of the integrated tensor, and do the non-relativistic approximation. \\
    We explain this procedure using $V_{F_{2,1}}$ as an example. In obtaining $V_{F_{2,1}}$, we need to calculate the following integral
    \begin{align}	    
        i\int\frac{\mu^{4-D}\d^Dl}{(2\pi)^D}\ \frac{\{1,l^\mu,l^\mu l^\rho\}}{(l^2-m^2+i\epsilon)[(l-q)^2-m^2+i\epsilon]}\notag
    \end{align}
    which equals to $\mathcal{J}^F_0$, $-q^\mu\mathcal{J}^F_{11}$ and $q^\mu q^\rho \mathcal{J}^F_{21}+g^{\mu\rho}{J}^F_{22}$ in the non-relativistic ChEFT, respectively. One can see that the coefficients of these three 
    tensors are
    \begin{align}
        0,\ 0,\ 4i(\epsilon_2\cdot\epsilon_4^*)(p_1+p_3)^\mu(p_2+p_4)^\nu\notag
    \end{align}
    After the contraction and under the non-relativistic approximation, we have 
    \begin{align}
        V_{F_{2,1}}^{\mathrm{R}}&=\frac{1}{2}\cdot\frac{3}{8f^4}\cdot 16m_Dm_{D^*}(\epsilon_2\cdot\epsilon_4^*)\mathcal{J}^F_{22}=V_{F_{2,1}}^{\mathrm{NR}}\cdot 4m_Dm_{D^*}. 
    \end{align}
    We also elaborate on $V_{T_{2,1}}^{\mathrm{R}}$ as an example. Our covariant potential has the following structure
    \begin{align}
        i\int\frac{\d^4l}{(2\pi)^4}\frac{\{1,l^{\mu},l^{\mu}l^{\nu},l^{\mu}l^{\nu}l^{\rho},\cdots\}}{[(p_2-l)^2-m_D^2+i\epsilon](l^2-m_\pi^2+i\epsilon)[(q+l)^2-m_\pi^2+i\epsilon]}.
    \end{align}
    According to the heavy meson approach, we deal with the first denominator by replacing $(p_2-l)_\mu=m_Dv_\mu+k_\mu$, and therefore have
    \begin{align}
        (p_2-l)^2-m_D^2\simeq2m_Dv\cdot k=2m_D(-v\cdot l+v\cdot p_2-m_D)
    \end{align}
    We rewrite $\omega=v\cdot p_2-m_D\simeq\Delta$, and under the non-relativistic approximation, we have
    \begin{align}
        &i\int\frac{\d^4l}{(2\pi)^4}\frac{\{1,l^{\mu},l^{\mu}l^{\nu},l^{\mu}l^{\nu}l^{\rho},\cdots\}}{[(p_2-l)^2-m_D^2+i\epsilon](l^2-m_\pi^2+i\epsilon)[(q+l)^2-m_\pi^2+i\epsilon]}\notag\\
        \simeq&\frac{i}{2m_D}\int\frac{\d^4l}{(2\pi)^4}\frac{\{1,l^{\mu},l^{\mu}l^{\nu},l^{\mu}l^{\nu}l^{\rho},\cdots\}}{(-v\cdot l+\omega+i\epsilon)(l^2-m_\pi^2+i\epsilon)[(l+q)^2-m_\pi^2+i\epsilon]}\notag\\
        \equiv&\frac{1}{2m_D}\{\J_{0}^{S},\ q^\mu\J_{11}^S+v^\mu\J_{12}^S,\ g^{\mu\nu}\J_{21}^S+q^\mu q^\nu\J_{22}^S+v^\mu v^\nu\J_{23}^S+(q\vee v)\J_{24}^S,\notag\\
        &(g\vee q)\J_{31}^{S}+q^\mu q^\nu q^\rho\J_{32}^{S}+(q^2\vee v)\J_{33}^{S}+(g\vee v)\J_{34}^{S}+(q\vee v^2)\J_{35}^{S}+v^\mu v^\nu v^\rho\J_{36}^{S},\ \cdots\}\notag\\
        \simeq&\frac{1}{2m_D}\{\J_{0}^{S},\ q^\mu\J_{11}^S+\frac{p_2^\mu}{m_{D^{*}}}\J_{12}^S,\ g^{\mu\nu}\J_{21}^S+q^\mu q^\nu\J_{22}^S+\frac{p_2^\mu p_2^\nu}{m_{D^{*}}^2}\J_{23}^S+\frac{q\vee p_2}{m_{D^*}}\J_{24}^S,\notag\\
        &(g\vee q)\J_{31}^{S}+q^\mu q^\nu q^\rho\J_{32}^{S}+\frac{q^2\vee p_2}{m_{D^*}}\J_{33}^{S}+\frac{g\vee p_2}{m_{D^*}}\J_{34}^{S}+\frac{q\vee p_2^2}{m_{D^*}^2}\J_{35}^{S}+\frac{p_2^\mu p_2^\nu p_2^\rho}{m_{D^{*}}^3}\J_{36}^{S},\ \cdots\}
    \end{align}
    Contract with the coefficients and do the non-relativistic approximation, we finally obtain
    \begin{align}
        V_{T_{2,1}}^{\mathrm{NR}}&=\frac{3g^2}{f^4}[(\epsilon_2\cdot\epsilon_4^*)\cdot \J_{34}^{S}+(q\cdot\epsilon_2)(q\cdot\epsilon_4^*)(\J_{24}^{S}+\J_{33}^{S})](m_\pi,\Delta,|\pmb{q}|)\cdot(-\frac{1}{4})\\
        V_{T_{2,1}}^{\mathrm{R}}&=\frac{3g_{D}^2}{2f^4}[-4m_Dm_{D^*}(q\cdot\epsilon_2)(q\cdot\epsilon_4^{*})\cdot\J_{24}^{S}-4m_Dm_{D^*}(q\cdot\epsilon_2)(q\cdot\epsilon_4^{*})\cdot\J_{33}^{S}-4m_Dm_{D^*}(\epsilon_2\cdot\epsilon_4^{*})\cdot\J_{34}^{S}]\frac{1}{2m_Dm_{D^*}}\notag\\
        &=V_{T_{2,1}}^{\mathrm{NR}}\cdot\frac{g_D^2}{g^2}\cdot4
    \end{align}
    
    Note that in our covariant potential, the order of the integrated tensor may be higher than what we have shown in Eqs.~\ref{2-point J func}--~\ref{4-point J func}. However, since $l^\mu l^\nu/m_{D^{(*)}}^2\simeq0$, the problem does not matter. 
\end{enumerate}

From the steps shown above, after dividing our potentials by $\sqrt{2m_D2m_{D^*}2m_D2m_{D^*}}$ according to Ref.~\cite{Yang:2011wz}, and note that $g_{D^*}\simeq g$, it is obvious that after taking the non-relativistic approximation, our covariant TPE potential is the same as the HMChEFT one.

\label{AppendixC}
\end{widetext}

\bibliography{Refs}

\end{document}